%% file: Draft_Vaibhav_v5.tex
\documentclass[twocolumn,conference]{IEEEtran}
\PassOptionsToPackage{ruled, linesnumbered}{algorithm2e}
\usepackage[T1]{fontenc}
\usepackage[latin9]{inputenc}
\usepackage{color}
\usepackage{array}
\usepackage{verbatim}
\usepackage{mathrsfs}
\usepackage{multirow}
\usepackage{algorithm2e}
\usepackage{amsmath}
\usepackage{amssymb}
\usepackage{graphicx}
\usepackage{tablefootnote}
\usepackage[unicode=true,
 bookmarks=true,bookmarksnumbered=true,bookmarksopen=true,bookmarksopenlevel=1,
 breaklinks=false,pdfborder={0 0 0},pdfborderstyle={},backref=false,colorlinks=false]
 {hyperref}
\hypersetup{pdftitle={Your Title},
 pdfauthor={Your Name},
 pdfpagelayout=OneColumn, pdfnewwindow=true, pdfstartview=XYZ, plainpages=false}

\makeatletter

\providecommand{\tabularnewline}{\\}

\usepackage{cite, amsmath, amsfonts, amssymb, graphicx,lipsum}
\usepackage{amsthm}
\usepackage{balance}

\theoremstyle{remark}

\newcommand{\herm}{^{{\dagger}}}
\newcommand{\trans}{^{\mathsf{T}}}

\DeclareMathOperator{\diag}{diag}
\DeclareMathOperator{\maximize}{maximize}
\DeclareMathOperator{\st}{subject~to}

\DeclareMathOperator{\argmax}{argmax}

\usepackage{pgfplots}
\usepackage{pgfplotstable}
\usepackage{tikz}
\usetikzlibrary{pgfplots.groupplots}

\makeatother

\begin{document}
\title{\huge{On the Secrecy Rate under Statistical QoS Provisioning \\for
RIS-Assisted MISO Wiretap Channel}}
\author{\IEEEauthorblockN{Vaibhav~Kumar\IEEEauthorrefmark{1}, Mark~F.~Flanagan\IEEEauthorrefmark{1},
Derrick~Wing~Kwan~Ng\IEEEauthorrefmark{2}, and Le-Nam~Tran\IEEEauthorrefmark{1}}\IEEEauthorblockA{\IEEEauthorrefmark{1}School of Electrical and Electronic Engineering,
University College Dublin, Belfield, Dublin 4, Ireland\\
\IEEEauthorrefmark{2}School of Electrical Engineering and Telecommunications,
University of New South Wales, NSW 2052, Australia \\
Email: \{vaibhav.kumar, mark.flanagan\}@ieee.org, w.k.ng@unsw.edu.au,
nam.tran@ucd.ie}}

\maketitle
{\let\thefootnote\relax\footnotetext{This publication has emanated from research conducted with the financial support of Science Foundation Ireland (SFI) and is co-funded under the European Regional Development Fund under Grant Number 17/CDA/4786.}}
\begin{abstract}
Reconfigurable intelligent surface (RIS) assisted radio is considered
as an enabling technology with great potential for the sixth-generation
(6G) wireless communications standard. The achievable secrecy rate
(ASR) is one of the most fundamental metrics to evaluate the capability
of facilitating secure communication for RIS-assisted systems. However,
the definition of ASR is based on Shannon's information theory, which
generally requires long codewords and thus fails to quantify the secrecy
of emerging delay-critical services. Motivated by this, in this paper
we investigate the problem of maximizing the secrecy rate under a
delay-limited quality-of-service (QoS) constraint, termed as the effective
secrecy rate (ESR), for an RIS-assisted multiple-input single-output
(MISO) wiretap channel subject to a transmit power constraint. We
propose an iterative method to find a stationary solution to the formulated
non-convex optimization problem using a block coordinate ascent method
(BCAM), where both the beamforming vector at the transmitter as well
as the phase shifts at the RIS are obtained in closed forms in each
iteration. We also present a convergence proof, an efficient implementation,
and the associated complexity analysis for the proposed method. Our
numerical results demonstrate that the proposed optimization algorithm
converges significantly faster that an existing solution.%
{} The simulation results also confirm that the secrecy rate performance
of the system with stringent delay requirements reduces significantly
compared to the system without any delay constraints, and that this
reduction can be significantly mitigated by an appropriately placed
large-size RIS.
\end{abstract}

\section{Introduction}

\allowdisplaybreaks \sloppy

With the recent development in programmable metasurface technology,
reconfigurable intelligent surfaces (RISs) is being considered as
a prominent candidate for the sixth-generation (6G) wireless communications
systems. In practice, an RIS consists of multiple low-cost passive
reflecting elements which are capable of steering the incident radio
waves in a desirable direction to optimize the system's performance~\cite{Zhang_ComMag}.
In addition to their ability to enhance communication metrics such
as the achievable rate, error rate, outage probability and energy
efficiency, RISs are also being considered as one of the most promising
candidates for physical-layer security (PLS) provisioning in the next-generation
communication systems~\cite{SecrecySurvey}.

To unlock the potential of RIS for secure communication, advanced
resource allocation has been studied under different scenarios. For
instance, in~\cite{MISO-RIS-AN}, the authors presented the problem
of maximizing the \textit{achievable} secrecy rate (ASR) for an RIS-assisted
multiple-input single-output (MISO-RIS) system with multiple eavesdroppers,
under a maximum transmit power budget. More specifically, the transmit
beamforming and RIS reflection coefficients were jointly optimized
using an alternating optimization (AO) based iterative algorithm and
it was concluded that RIS helps to enhance the secrecy of the MISO
system. %
Also, in~\cite{NoEveCSI}, the problem of ASR maximization for a
MISO-RIS system without the eavesdropper's channel state information
(CSI) was studied. %
To obtain a suboptimal solution of the underlying non-convex optimization
problem, the authors in~\cite{NoEveCSI} adopted oblique manifold
and minorization-maximization algorithms, where the former was shown
to offer a higher ASR for a large number of reflecting elements in
the high-SNR regime.

On the other hand, the problem of ASR maximization for a MISO-RIS
millimeter wave (mmWave) system with multiple (colluding and non-colluding)
eavesdroppers with imperfect CSI at the transmitter was addressed
using AO and semidefinite relaxation (SDR) in~\cite{mmWaveMISO}.
Besides, the problem of ASR maximization for an RIS-assisted multiple-input
multiple-output (MIMO) downlink system was studied in~\cite{MIMO},
where an AO-based iterative scheme was used to jointly optimize the
transmit covariance matrix (at the transmitter) and the phase shifts
(at the RIS). Furthermore, a bisection-search-based AO scheme was
applied in~\cite{AO-Bisection} to maximize the ASR for a MISO-RIS
system. In order to jointly optimize the transmit beamforming vector
and the phase shifts%
{} in~\cite{AO-Bisection}, a closed-form expression for the transmitter
structure was first obtained for given phase shifts, and then the
phase shifts were obtained using a bisection search for a given beamforming
vector.

Despite the fruitful research in the literature, the definition of
ASR in those works, e.g.,~\cite{MISO-RIS-AN,NoEveCSI,mmWaveMISO,MIMO,AO-Bisection},
is based on the Shannon's definition of achievable rate with infinitely
long channel codes that does not account for the delay requirements
of the legitimate receiver(s). However, 6G communications standard
will be expected to support numerous delay-sensitive applications
including autonomous driving, unmanned aerial vehicles (UAVs), tactile
Internet, ultra-high-definition live video streaming, and critical
healthcare and military services, etc. In order to quantify the maximum
constant arrival rate of a service process with guaranteed statistical
delay QoS provisioning, the notion of \textit{effective rate} (ER)
was introduced in~\cite{EffectiveRate}. Based on this result, the
authors in~\cite{ESR} defined the novel concept of \textit{effective
secrecy rate} (ESR) as the maximum constant arrival rate that can
be supported securely over a radio link while satisfying the delay
QoS requirement, where the secrecy rate was considered as the service
rate.%
{} However, to the best of the authors' knowledge, the ESR for an RIS-assisted
systems has not yet been investigated in the literature\textcolor{blue}{{}
}and the existing results are not applicable to the problem in interest.
Therefore, in this paper, we study the ESR maximization of MISO-RIS
system, with main contributions listed below:
\begin{itemize}
\item We present the secrecy rate analysis under statistical QoS provisioning
of an RIS-assisted system, where the transmitter is equipped with
multiple antennas, while the legitimate receiver and the eavesdropper
are each equipped with a single antenna. In particular, by assuming
that the instantaneous CSI for all of the wireless links are available
at all of the nodes, we formulate an ESR maximization problem and
then propose a closed-form-solution-based block coordinate ascent
method (BCAM) to find a stationary solution. By Monte-Carlo simulations,
we show that the proposed algorithm converges faster than the existing
bisection-search-based AO algorithm.
\item \textcolor{black}{We prove the convergence and provide an efficient
implementation and the associated complexity analysis of the proposed
algorithm.}
\item Finally, we perform extensive numerical experiments to show the impacts
of different system parameters including the delay QoS exponent, the
number of transmit antennas, and the number of reflecting elements
in the RIS on the ESR of the considered MISO-RIS system.
\end{itemize}

\paragraph*{Notations}

We use bold uppercase and lowercase letters to denote matrices and
vectors, respectively. $\left\Vert \mathbf{X}\right\Vert ,\ \mathbf{X^{*},}\ \mathbf{X}\trans$,
and $\mathbf{X}\herm$ denote the Frobenius norm, complex conjugate,
(ordinary) transpose, and Hermitian transpose of $\mathbf{X}$, respectively,
while, $\left|x\right|$ denotes the absolute value of the complex
number $x$. We denote the space of complex matrices of size $M\times N$
by $\mathbb{C}^{M\times N}$, $\mathbb{E}\left\{ \cdot\right\} $
denotes the expectation operator,\textcolor{blue}{{} }and $\Re\left\{ \cdot\right\} $
denotes the real part of a complex number. $\diag\left(\mathbf{x}\right)$
denotes the square diagonal matrix which has the elements of $\mathbf{x}$
on the main diagonal, and $\mathbf{I}$ denotes the identity matrix.

\section{System Model and Problem Formulation}

\begin{figure}[t]
\begin{centering}
\includegraphics[width=0.9\columnwidth]{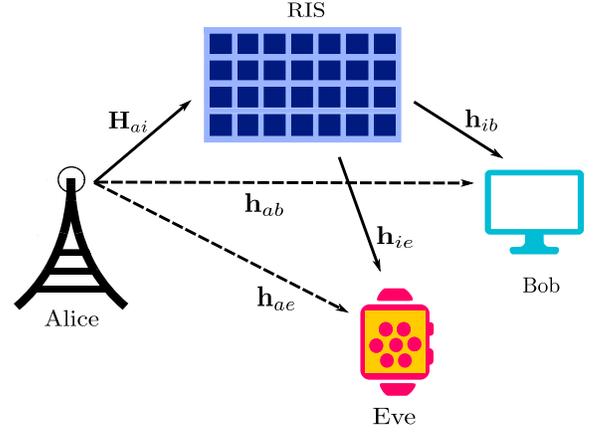}
\par\end{centering}
\caption{System model for RIS-assisted MISO wiretap channel.}
\label{Fig:SysMod}
\end{figure}

Consider the MISO-RIS system shown in Fig.~\ref{Fig:SysMod}, consisting
of a transmitter (Alice), one legitimate receiver (Bob), one eavesdropper
(Eve), and one RIS. It is assumed that Alice is equipped with $N\geq1$
antennas, while Bob and Eve are single-antenna devices. The RIS is
assumed to have $N_{\mathrm{ris}}$ passive reflecting elements. The
channel matrix between Alice and RIS is denoted by $\mathbf{H}_{ai}\in\mathbb{C}^{N_{\mathrm{ris}}\times N}$,
while the channel vectors of Alice-Bob and Alice-Eve links are denoted
by $\mathbf{h}_{ab}\in\mathbb{C}^{1\times N}$ and $\mathbf{h}_{ae}\in\mathbb{C}^{1\times N}$,
respectively. The channel vectors of the RIS-Bob and RIS-Eve links
are denoted by $\mathbf{h}_{ib}\in\mathbb{C}^{1\times N_{\mathrm{ris}}}$
and $\mathbf{h}_{ie}\in\mathbb{C}^{1\times N_{\mathrm{ris}}}$, respectively.
Considering that Alice transmits a secret message~$s\in\mathbb{C}$
intended for Bob, the signals received at Bob and Eve are, respectively,
given by 
\begin{equation}
\begin{aligned}y_{b} & =\mathbf{h}_{ib}\diag\left(\boldsymbol{\theta}\right)\mathbf{H}_{ai}\mathbf{w}s+\mathbf{h}_{ab}\mathbf{w}s+n_{b},\\
y_{e} & =\mathbf{h}_{ie}\diag\left(\boldsymbol{\theta}\right)\mathbf{H}_{ai}\mathbf{w}s+\mathbf{h}_{ae}\mathbf{w}s+n_{e},
\end{aligned}
\label{eq:RxSignal}
\end{equation}
where $\mathbf{w}\in\mathbb{C}^{N\times1}$ is the transmit beamforming
vector, $\boldsymbol{\theta}\triangleq\left[\theta_{1}\ \theta_{2}\ \cdots\ \theta_{N_{\mathrm{ris}}}\right]\trans\in\mathbb{C}^{N_{\mathrm{ris}}\times1}$,
$\theta_{l}=\exp\left(j\phi_{l}\right)$, $\phi_{l}\in\left[0,2\pi\right)$
denotes the phase shift induced by the $l$-th reflecting element
in the RIS with $l\in\mathcal{L}$, $\mathcal{L}=\left\{ 1,2,\ldots,N_{\mathrm{ris}}\right\} $,
and $n_{b}\sim\mathcal{CN}\left(0,\sigma_{b}^{2}\right)$ and $n_{e}\sim\mathcal{CN}\left(0,\sigma_{e}^{2}\right)$
denote the additive white Gaussian noise at Bob and Eve, respectively.
Here we have $\left\Vert \mathbf{w}\right\Vert ^{2}\leq P$, where
$P$ denotes the maximum transmit power budget at Alice. Assuming
that instantaneous CSI of all the wireless links are available at
all nodes as well as the RIS\footnote{A similar assumption was considered in many papers, including~\cite{AO-Bisection,MIMO}.
Such a scenario is possible where Eve is also a legitimate receiver
for Alice, but is untrusted for Bob. However, the scenario where only
partial or no CSI is available at Alice will be considered in a future
work.}, the ASR for Bob (in bps/Hz), for given $\mathbf{w}$ and $\boldsymbol{\theta}$,
is given by 
\begin{equation}
R_{s}\left(\mathbf{w},\boldsymbol{\theta}\right)=\log_{2}\left(\dfrac{1+\left|\mathbf{z}_{b}\left(\boldsymbol{\theta}\right)\mathbf{w}\right|^{2}}{1+\left|\mathbf{z}_{e}\left(\boldsymbol{\theta}\right)\mathbf{w}\right|^{2}}\right),\label{eq:ASR_Def}
\end{equation}
where $\mathbf{z}_{b}\left(\boldsymbol{\theta}\right)\triangleq\hat{\mathbf{h}}_{ib}\diag\left(\boldsymbol{\theta}\right)\mathbf{H}_{ai}+\hat{\mathbf{h}}_{ab}$,
$\mathbf{z}_{e}\triangleq\hat{\mathbf{h}}_{ie}\diag\left(\boldsymbol{\theta}\right)\mathbf{H}_{ai}+\hat{\mathbf{h}}_{ae}$,
$\hat{\mathbf{h}}_{ib}=\mathbf{h}_{ib}/\sigma_{b}$, $\hat{\mathbf{h}}_{ab}=\mathbf{h}_{ab}/\sigma_{b}$,
$\hat{\mathbf{h}}_{ie}=\mathbf{h}_{ie}/\sigma_{e}$, and $\hat{\mathbf{h}}_{ae}=\mathbf{h}_{ae}/\sigma_{e}$.

{} We remark that the ASR is a useful performance metric for delay-tolerant
applications. However, in this paper we are interested in the effective
secrecy rate where some degree of QoS constraints must be satisfied.
Let $\vartheta$ be the rate at which the buffer occupancy at Alice
decays, where $\vartheta\triangleq-\lim_{x\to\infty}\tfrac{1}{x}\Pr\left\{ \mathscr{L}>x\right\} $,
with $\mathscr{L}$ being the queue length at equilibrium. Also, let
$T$ denote the coherence time of all of the wireless links which
is an integer multiple of the symbol duration and $\mathcal{B}$ denote
the total available bandwidth. Since the channels are known to all
the involving nodes, Alice can adapt the wiretap coding with respective
to each channel realization to maximize the secrecy service rate.
In this regard, the ESR for Bob (in bps/Hz) is defined as~(c.f.~\cite[eqn. (12)]{ESR,MIMO_Queue})
\begin{multline}
E_{s}\left(\mathbf{w},\boldsymbol{\theta}\right)\\
\triangleq\!\dfrac{-1}{\vartheta T\mathcal{B}}\ln\!\left(\!\mathbb{E}\left\{ \exp\left(\!-\vartheta T\mathcal{B}\underset{\mathbf{w}\in\mathcal{S}_{w},\boldsymbol{\theta}\in\mathcal{S}_{\theta}}{\mathrm{max}\ }R_{s}\left(\mathbf{w},\boldsymbol{\theta}\right)\!\right)\!\right\} \!\right),\label{eq:ESR_Def}
\end{multline}
where the expectation is performed over the involved channels, and
$\mathcal{S}_{w}$ and $\mathcal{S}_{\theta}$ are defined as 
\begin{equation}
\begin{aligned}\mathcal{S}_{w} & \triangleq\left\{ \mathbf{w}\left|\left\Vert \mathbf{w}\right\Vert ^{2}\leq P\right.\right\} ,\\
\mathcal{S}_{\theta} & \triangleq\left\{ \boldsymbol{\theta}\left|\left|\theta_{l}\right|=1,l\in\mathcal{L}\right.\right\} .
\end{aligned}
\label{eq:FeasibleSets-1}
\end{equation}
Note that $\vartheta\to0$ represents the system without any delay
constraint, whereas, $\vartheta\to\infty$ corresponds to the system
with extremely-strict delay constraint. For the case when $\vartheta\to0$,
the ESR is identical to the ASR.  It is obvious that to evaluate
the effective secrecy rate, we need to solve the following optimization
problem:
\begin{equation}
\begin{aligned}\underset{\mathbf{w},\boldsymbol{\theta}}{\maximize\ } & f\left(\mathbf{w},\boldsymbol{\theta}\right)\\
\st\  & \mathbf{w}\in\mathcal{S}_{w},\ \boldsymbol{\theta}\in\mathcal{S}_{\theta}.
\end{aligned}
\label{eq:OptProbFinal}
\end{equation}
where 
\begin{equation}
f\left(\mathbf{w},\boldsymbol{\theta}\right)\triangleq\dfrac{1+\left|\mathbf{z}_{b}\left(\boldsymbol{\theta}\right)\mathbf{w}\right|^{2}}{1+\left|\mathbf{z}_{e}\left(\boldsymbol{\theta}\right)\mathbf{w}\right|^{2}}.\label{eq:fDef}
\end{equation}
It is not surprising that when channels are perfectly known at all
nodes, finding the effective secrecy rate boils down to solving the
conventional secrecy rate maximization problem. In the following section,
we propose a method based on the concept of block coordinate ascent
method to maximize the objective.

\section{Proposed Solution based on Block Coordinate Ascent Method}

In this section, we present a computationally efficient algorithm
to find a stationary solution to the problem in~\eqref{eq:OptProbFinal}
that optimizes the beamforming vector $\mathbf{w}$ and each individual
phase shift of the RIS using the BCAM. More specifically, we optimize
$\mathbf{w}$ for a given $\boldsymbol{\theta}$, and optimize a specific
phase shift $\theta_{l}$ for when $\mathbf{w}$ and other phase shifts
$\theta_{m\neq l}$ are held fixed. These two steps are achieved in
closed-forms as described in the next two subsections.

\subsection{Closed-Form Expression for Optimal $\mathbf{w}$ for Given $\boldsymbol{\theta}$}

For a given phase shift vector $\boldsymbol{\theta},$ the optimization
over $\mathbf{w}$ is expressed as
\begin{equation}
\begin{aligned}\maximize\  & \dfrac{1+\mathbf{w}\herm\mathbf{Z}_{b}\left(\boldsymbol{\theta}\right)\mathbf{w}}{1+\mathbf{w}\herm\mathbf{Z}_{e}\left(\boldsymbol{\theta}\right)\mathbf{w}}\\
\st\  & \mathbf{w}\in\mathcal{S}_{w},
\end{aligned}
\label{eq:OptProbGivenTheta}
\end{equation}
where $\mathbf{Z}_{b}\left(\boldsymbol{\theta}\right)\triangleq\mathbf{z}_{b}\herm\left(\boldsymbol{\theta}\right)\mathbf{z}_{b}\left(\boldsymbol{\theta}\right)$
and $\mathbf{Z}_{e}\left(\boldsymbol{\theta}\right)\triangleq\mathbf{z}_{e}\herm\left(\boldsymbol{\theta}\right)\mathbf{z}_{e}\left(\boldsymbol{\theta}\right)$,
which admits a closed-form solution given by~(c.f.~\cite{Khisti})
\begin{equation}
\mathbf{w}_{\mathrm{opt}}=\sqrt{P}\mathbf{u}_{\max},\label{eq:W_Opt}
\end{equation}
where $\mathbf{u}_{\max}$ is the normalized eigenvector associated
with the maximum eigenvalue of the matrix $\left(P\mathbf{Z}_{e}\left(\boldsymbol{\theta}\right)+\mathbf{I}\right)^{-1}\left(P\mathbf{Z}_{b}\left(\boldsymbol{\theta}\right)+\mathbf{I}\right)$.

\subsection{Closed-Form Expression for Optimal $\theta_{l}$ for Given $\mathbf{w}$
and other $\theta_{m}\left(m\protect\neq l\right)$}

In this subsection, we derive a closed-form expression for optimal
$\theta_{l}$ while other variables (including $\mathbf{w}$ and $\theta_{m}$,
$m\in\left\{ \mathcal{L}\setminus l\right\} $) are kept fixed. In
fact, there are a few existing methods to find the phase shifts for
a given $\mathbf{w}$. In~\cite{Zhang}, a combination of semi-definite
rank relaxation method and Gaussian randomization was used.%
{} Yet, such a method incurs high complexity since a semi-definite program
needs to be solved in each iteration. In~\cite{AO-Bisection}, the
authors applied Dinkelbach's method together with the majorization-minimization
principle to maximize a lower bound of $f\left(\mathbf{w},\boldsymbol{\theta}\right)$
in each iteration. The main drawback of this method is that a bisection
search is still required, each iteration of which involves computing
the maximum eigenvalue of a large matrix whose dimension is the number
of reflecting elements (this value can be in the order of hundreds
or even thousands in practically-envisioned RIS deployments).

Different from the existing solutions, our method can be viewed as
a generalization of~\cite{Schober_MISO}. Note that the direct links
for Alice-Bob and Alice-Eve channels were not considered in~\cite{Schober_MISO},
and therefore it is not straightforward to adopt the solution proposed
in~\cite{Schober_MISO}. Using the relation $\mathbf{x}\diag\left(\boldsymbol{\theta}\right)=\boldsymbol{\theta}\trans\diag\left(\mathbf{x}\right)$,
we can further express the numerator of~\eqref{eq:OptProbGivenTheta}
as%
\begin{align}
 & 1+\mathbf{w}\herm\mathbf{Z}_{b}\left(\boldsymbol{\theta}\right)\mathbf{w}\nonumber \\
\!\!\mathcal{}\!=\  & 1\!+\!\mathbf{w}\herm\!\left[\hat{\mathbf{h}}_{ib}\diag\left(\boldsymbol{\theta}\right)\mathbf{H}_{ai}\!+\!\hat{\mathbf{h}}_{ab}\right]\herm\!\left[\hat{\mathbf{h}}_{ib}\mathrm{\diag\left(\boldsymbol{\theta}\right)}\mathbf{H}_{ai}\!+\!\hat{\mathbf{h}}_{ab}\right]\mathbf{w}\nonumber \\
\!\!\!\triangleq\  & 1+\boldsymbol{\theta}\herm\mathbf{A}_{b}\left(\mathbf{w}\right)\boldsymbol{\theta}+\mathbf{b}_{b}\herm\left(\mathbf{w}\right)\boldsymbol{\theta}+\boldsymbol{\theta}\herm\mathbf{b}_{b}\left(\mathbf{w}\right)+c_{b}\left(\mathbf{w}\right),\label{eq:NumSimplified}
\end{align}
where 
\begin{equation}
\begin{aligned} & \!\!\!\!\mathbf{A}_{b}\left(\mathbf{w}\right)\!=\!\mathbf{a}_{b}\left(\mathbf{w}\right)\mathbf{a}_{b}\herm\left(\mathbf{w}\right);\ \mathbf{\mathbf{a}}_{b}\left(\mathbf{w}\right)\!=\!\diag\left(\hat{\mathbf{h}}_{ib}^{*}\right)\mathbf{H}_{ai}^{*}\mathbf{w}^{*};\\
 & \!\!\!\!\mathbf{b}_{b}\left(\mathbf{w}\right)\!=\!\diag\left(\hat{\mathbf{h}}_{ib}^{*}\right)\mathbf{H}_{ai}^{*}\mathbf{w}^{*}\mathbf{w}\trans\hat{\mathbf{h}}_{ab}\trans;\ c_{b}\left(\mathbf{w}\right)\!=\!\left|\hat{\mathbf{h}}_{ab}\mathbf{w}\right|^{2}.
\end{aligned}
\label{eq:NumCompDef}
\end{equation}
 Following a similar set of arguments, the denominator of~\eqref{eq:OptProbGivenTheta}
can be represented as 
\begin{align}
 & 1+\mathbf{w}\herm\mathbf{Z}_{e}\left(\boldsymbol{\theta}\right)\mathbf{w}\nonumber \\
=\  & 1+\boldsymbol{\theta}\herm\mathbf{A}_{e}\left(\mathbf{w}\right)\boldsymbol{\theta}+\mathbf{b}_{e}\herm\left(\mathbf{w}\right)\boldsymbol{\theta}+\boldsymbol{\theta}\herm\mathbf{b}_{e}\left(\mathbf{w}\right)+c_{e}\left(\mathbf{w}\right),\label{eq:DenSimplified}
\end{align}
where

\begin{equation}
\begin{aligned} & \!\!\!\!\mathbf{A}_{e}\left(\mathbf{w}\right)\!=\!\mathbf{a}_{e}\left(\mathbf{w}\right)\mathbf{a}_{e}\herm\left(\mathbf{w}\right);\ \mathbf{\mathbf{a}}_{e}\left(\mathbf{w}\right)\!=\!\diag\left(\hat{\mathbf{h}}_{ie}^{*}\right)\mathbf{H}_{ai}^{*}\mathbf{w}^{*};\\
 & \!\!\!\!\mathbf{b}_{e}\left(\mathbf{w}\right)\!=\!\diag\left(\hat{\mathbf{h}}_{ie}^{*}\right)\mathbf{H}_{ai}^{*}\mathbf{w}^{*}\mathbf{w}\trans\hat{\mathbf{h}}_{ae}\trans;\ c_{e}\left(\mathbf{w}\right)\!=\!\left|\hat{\mathbf{h}}_{ae}\mathbf{w}\right|^{2}.
\end{aligned}
\label{eq:DenCompDef}
\end{equation}
Therefore, the optimization problem in~\eqref{eq:OptProbFinal} for
a given $\mathbf{w}$ can be given by 
\begin{equation}
\begin{aligned}\underset{\boldsymbol{\theta}}{\maximize\ } & \dfrac{1+\boldsymbol{\theta}\herm\mathbf{A}_{b}\left(\mathbf{w}\right)\boldsymbol{\theta}+\mathbf{b}_{b}\herm\left(\mathbf{w}\right)\boldsymbol{\theta}+\boldsymbol{\theta}\herm\mathbf{b}_{b}\left(\mathbf{w}\right)+c_{b}\left(\mathbf{w}\right)}{1+\boldsymbol{\theta}\herm\mathbf{A}_{e}\left(\mathbf{w}\right)\boldsymbol{\theta}+\mathbf{b}_{e}\herm\left(\mathbf{w}\right)\boldsymbol{\theta}+\boldsymbol{\theta}\herm\mathbf{b}_{e}\left(\mathbf{w}\right)+c_{e}\left(\mathbf{w}\right)}\\
\st\  & \boldsymbol{\theta}\in\mathcal{S}_{\theta}.
\end{aligned}
\label{eq:OptProbGivenW}
\end{equation}

To realize a more efficient method, we sequentially optimize each
$\theta_{l}$ at a time while the other phase shifts (along with the
other variables) are held fixed. To lighten the notations, we write
$\mathbf{A}_{b}$ instead of $\mathbf{A}_{b}\left(\mathbf{w}\right)$
onward. The same applies to other quantities in~\eqref{eq:NumCompDef}
and the quantities in~\eqref{eq:DenCompDef}. Let $\mathbf{a}_{b}=\left[a_{b1}\ a_{b2}\ \cdots\ a_{bN_{\mathrm{ris}}}\right]\trans$,
$\mathbf{a}_{e}=\left[a_{e1}\ a_{e2}\ \cdots\ a_{eN_{\mathrm{ris}}}\right]\trans$,
$\mathbf{b}_{b}=\left[b_{b1}\ b_{b2}\ \cdots\ b_{bN_{\mathrm{ris}}}\right]\trans$
and $\mathbf{b}_{e}=\left[b_{e1}\ b_{e2}\ \cdots\ b_{eN_{\mathrm{ris}}}\right]\trans$.
Then the maximization over a specific $\theta_{l}$ is expressed as
\begin{equation}
\begin{aligned}\underset{\theta_{l}}{\maximize}\  & \dfrac{\Re\left\{ \alpha_{bl}^{*}\theta_{l}\right\} +\beta_{bl}}{\Re\left\{ \alpha_{el}^{*}\theta_{l}\right\} +\beta_{el}}\\
\st\  & \left|\theta_{l}\right|=1,
\end{aligned}
\label{eq:OptProbGivenW-I}
\end{equation}
where $\alpha_{bl}=2\left(a_{bl}\sum\nolimits _{m\in\left\{ \mathcal{L}\setminus l\right\} }a_{bm}^{*}\theta_{m}+b_{bl}\right),$
$\beta_{bl}=\left|a_{bl}\right|^{2}+\left|\sum\nolimits _{m\in\left\{ \mathcal{L}\setminus l\right\} }a_{bm}^{\ast}\theta_{m}\right|^{2}+2\Re\left\{ \sum\nolimits _{m\in\left\{ \mathcal{L}\setminus l\right\} }b_{bm}^{\ast}\theta_{m}\right\} +c_{b}+1$,
$\alpha_{el}=2\left(a_{el}\sum\nolimits _{m\in\left\{ \mathcal{L}\setminus l\right\} }a_{em}^{*}\theta_{m}+b_{el}\right)$
and $\beta_{el}=\left|a_{el}\right|^{2}+\left|\sum\nolimits _{m\in\left\{ \mathcal{L}\setminus l\right\} }a_{em}^{\ast}\theta_{m}\right|^{2}+2\Re\left\{ \sum\nolimits _{m\in\left\{ \mathcal{L}\setminus l\right\} }b_{em}^{\ast}\theta_{m}\right\} +c_{e}+1$.
To proceed further, we rewrite $\alpha_{bl}=r_{bl}\exp\left(j\phi_{bl}\right)$,
$\alpha_{el}=r_{el}\exp\left(j\phi_{el}\right)$ and $\theta_{l}=\exp\left(j\phi_{l}\right)$.
Then,~\eqref{eq:OptProbGivenW-I} reduces to
\begin{equation}
\begin{aligned}\underset{\phi_{l}}{\maximize}\  & \dfrac{r_{bl}\cos\left(\phi_{l}-\phi_{bl}\right)+\beta_{bl}}{r_{el}\cos\left(\phi_{l}-\phi_{el}\right)+\beta_{el}}\triangleq g\left(\phi_{l}\right)\\
\st\  & 0\leq\phi_{l}<2\pi.
\end{aligned}
\label{eq:OptProbGivenW-II}
\end{equation}
Differentiating the objective function in~\eqref{eq:OptProbGivenW-II}
w.r.t. $\phi_{l}$, we get 
\begin{align}
 & g'\left(\phi_{l}\right)=\dfrac{-r_{bl}\sin\left(\phi_{l}-\phi_{bl}\right)\left\{ r_{el}\cos\left(\phi_{l}-\phi_{el}\right)+\beta_{el}\right\} }{\left\{ r_{el}\cos\left(\phi_{l}-\phi_{el}\right)+\beta_{el}\right\} ^{2}}\nonumber \\
 & \qquad\qquad+\dfrac{r_{el}\sin\left(\phi_{l}-\phi_{el}\right)\left\{ r_{bl}\cos\left(\phi_{l}-\phi_{bl}\right)+\beta_{bl}\right\} }{\left\{ r_{el}\cos\left(\phi_{l}-\phi_{el}\right)+\beta_{el}\right\} ^{2}}\nonumber \\
 & =\!\frac{r_{bl}r_{el}\sin\left(\!\phi_{bl}\!-\!\phi_{el}\!\right)\!+\!r_{el}\beta_{bl}\sin\left(\!\phi_{l}\!-\!\phi_{el}\!\right)\!-\!r_{bl}\beta_{el}\sin\left(\!\phi_{l}\!-\!\phi_{bl}\!\right)}{\left\{ r_{el}\cos\left(\phi_{l}-\phi_{el}\right)+\beta_{el}\right\} ^{2}}.\label{eq:GDiff-I}
\end{align}
Using the following relation, 
\[
r_{bl}\beta_{el}\sin\left(\phi_{l}-\phi_{bl}\right)-r_{el}\beta_{bl}\sin\left(\phi_{l}-\phi_{el}\right)=r_{l}\sin\left(\phi_{l}+\varphi_{l}\right),
\]
where
\begin{align*}
r_{l}= & \sqrt{r_{bl}^{2}\beta_{el}^{2}+r_{el}^{2}\beta_{bl}^{2}-2r_{bl}r_{el}\beta_{bl}\beta_{el}\cos\left(\phi_{el}-\phi_{bl}\right)},\\
\varphi_{l}= & \arctan\left[\dfrac{-r_{bl}\beta_{el}\sin\left(\phi_{bl}\right)+r_{el}\beta_{bl}\sin\left(\phi_{el}\right)}{r_{bl}\beta_{el}\cos\left(\phi_{bl}\right)-r_{el}\beta_{bl}\cos\left(\phi_{el}\right)}\right],
\end{align*}
we can rewrite $g'\left(\phi_{l}\right)$ as
\begin{equation}
g'\left(\phi_{l}\right)=\dfrac{r_{bl}r_{el}\sin\left(\phi_{bl}-\phi_{el}\right)-r_{l}\sin\left(\phi_{l}+\varphi_{l}\right)}{\left\{ r_{el}\cos\left(\phi_{l}-\phi_{el}\right)+\beta_{el}\right\} ^{2}}.\label{eq:GDiff-II}
\end{equation}
Note that if $\left|r_{bl}r_{el}\sin\left(\phi_{bl}-\phi_{el}\right)\right|>r_{l}$,
then $g'\left(\phi_{l}\right)$ is either positive or negative (i.e.,
$g'\left(\phi_{l}\right)$ can not be equal to zero) for all $\phi_{l}\in\left[0,2\pi\right)$.
As a result, $g(\phi_{l})$ is maximized when $\phi_{l}=0$, resulting
in $\theta_{l}=1$. On the other hand, if $\left|r_{bl}r_{el}\sin\left(\phi_{bl}-\phi_{el}\right)\right|\leq r_{l}$,
then $g'\left(\phi_{l}\right)=0$ has two solutions: 
\begin{equation}
\begin{aligned}\phi_{l_{1}} & =\arcsin\left[\tfrac{r_{bl}r_{el}}{r_{l}}\sin\left(\phi_{bl}-\phi_{el}\right)\right]-\varphi_{l},\\
\phi_{l_{2}} & =\pi-\arcsin\left[\tfrac{r_{bl}r_{el}}{r_{l}}\sin\left(\phi_{bl}-\phi_{el}\right)\right]-\varphi_{l}.
\end{aligned}
\label{eq:phi_l}
\end{equation}
Thus the optimal solution to~\eqref{eq:OptProbGivenW-I} admits a
closed-form expression given by 
\begin{equation}
\theta_{l,\mathrm{opt}}=\exp\left(j\phi_{l,\mathrm{opt}}\right),\label{eq:Theta_l_Opt}
\end{equation}
where 
\begin{equation}
\phi_{l,\mathrm{opt}}=\argmax\left\{ g\left(0\right),g\left(\phi_{l_{1}}\right),g\left(\phi_{l_{2}}\right)\right\} ,\label{eq:phi_l_Opt}
\end{equation}
and $\phi_{l_{1}}$ and $\phi_{l_{2}}$ are given in~\eqref{eq:phi_l}.
The overall algorithm is summarized in~Algorithm~\ref{alg:AO}.

\begin{algorithm}[t]
\caption{Block Coordinate Ascent Method}

\label{alg:AO}

\KwIn{ $\boldsymbol{\theta}_{0}$, $\mathbf{w}_{0}$ }

\KwOut{ $\boldsymbol{\theta}_{n},\mathbf{w}_{n}$}

$n\leftarrow1$\;

\Repeat{convergence }{

Given $\boldsymbol{\theta}_{n-1}$, compute $\mathbf{w}_{n}$ via
\eqref{eq:W_Opt}

\For{$l\in\mathcal{L}$}{

Compute $\theta_{l,\mathrm{opt}}$ for given $\mathbf{w}_{n}$ via
\eqref{eq:Theta_l_Opt}

$\theta_{l}\leftarrow\theta_{l,\mathrm{opt}}$

}

$n\leftarrow n+1$

}
\end{algorithm}

\subsection{Convergence Analysis}

We now show that the iterates generated by Algorithm~\ref{alg:AO}
converge to a stationary solution of~\eqref{eq:OptProbFinal}. First,
it is easy to see that $f(\mathbf{w}_{n+1},\boldsymbol{\theta}_{n+1})\geq f(\mathbf{w}_{n+1},\boldsymbol{\theta}_{n})\geq f(\mathbf{w}_{n},\boldsymbol{\theta}_{n})$
and thus Algorithm~\ref{alg:AO} generates a non-decreasing objective
sequence. It is also trivial to check that the objective function
is continuous and bounded from above. Moreover, the feasible set is
compact. Thus, the objective sequence must converge to certain limit,
i.e., $\underset{n\to\infty}{\lim}f(\mathbf{w}_{n},\boldsymbol{\theta}_{n})=f_{\ast}$.
Let $\mathcal{S}=\{(\mathbf{w},\boldsymbol{\theta})\ |\ f(\mathbf{w},\boldsymbol{\theta})\leq f_{\ast}\}$.
By the continuity of $f(\mathbf{w},\boldsymbol{\theta})$ and the
compactness of the feasible set, it is obvious that $\mathcal{S}$
is compact. Thus there exists a subsequence $(\mathbf{w}_{n_{j}},\boldsymbol{\theta}_{n_{j}})$
converging to the limit point $(\mathbf{w}_{\ast},\boldsymbol{\theta}_{\ast})$.
By continuity of $f(\mathbf{w},\boldsymbol{\theta})$ we must have
$f_{\ast}=f(\mathbf{w}_{\ast},\boldsymbol{\theta}_{\ast})$. The proof
that $(\mathbf{w}_{\ast},\boldsymbol{\theta}_{\ast})$ is a stationary
solution of \eqref{eq:OptProbFinal} is rather standard and thus is
omitted here for the sake of brevity. We refer the interested reader
to~\cite[Sec. 2.7]{Bertsekas} for further details.

\subsection{Efficient Implementation and Complexity Analysis}

We now provide an efficient implementation and the associated complexity
analysis of the proposed solution. For the complexity analysis, we
use the conventional big-O notation and focus on the number of complex
multiplications. First, note that $P\mathbf{Z}_{b}\left(\boldsymbol{\theta}\right)$
can be represented as $\left(P\mathbf{z}_{b}\left(\boldsymbol{\theta}\right)\right)\herm\mathbf{z}_{b}\left(\boldsymbol{\theta}\right)$.
The number of multiplications required to compute $P\mathbf{z}_{b}\left(\boldsymbol{\theta}\right)$
is equal to $\mathcal{O}\left(NN_{\mathrm{ris}}\right)$ and the number
of multiplications required to compute $P\mathbf{Z}_{b}\left(\boldsymbol{\theta}\right)$
is given by $N^{2}/2$. Note that the $\mathbf{w}$-update requires
the eigenvector associated with the maximum eigenvalue of $\bigl(P\mathbf{Z}_{e}\left(\boldsymbol{\theta}\right)+\mathbf{I}\bigr)^{-1}\bigl(P\mathbf{Z}_{b}\left(\boldsymbol{\theta}\right)+\mathbf{I}\bigr)$.
If we compute this term in a straightforward manner, it would take
$\mathcal{O}\left(N^{3}\right)$ complex multiplications. We now provide
an efficient way to achieve this, which has not been discussed in
the related literature. Using the Woodbury matrix identity~\cite{Woodbury},
it can be shown that $P\mathbf{Z}_{e}\left(\boldsymbol{\theta}\right)+\mathbf{I}\bigr)^{-1}=\mathbf{I}-\tfrac{P\mathbf{z}_{e}\left(\boldsymbol{\theta}\right)\herm\mathbf{z}_{e}\left(\boldsymbol{\theta}\right)}{1+P\mathbf{z}_{e}\left(\boldsymbol{\theta}\right)\mathbf{z}_{e}\herm\left(\boldsymbol{\theta}\right)}$.
Therefore, we have 
\begin{align*}
 & \bigl(P\mathbf{Z}_{e}\left(\boldsymbol{\theta}\right)+\mathbf{I}\bigr)^{-1}\bigl(P\mathbf{Z}_{b}\left(\boldsymbol{\theta}\right)+\mathbf{I}\bigr)\\
=\  & \mathbf{I}+P\mathbf{Z}_{b}\left(\boldsymbol{\theta}\right)-\dfrac{P^{2}\mathbf{z}_{e}\left(\boldsymbol{\theta}\right)\herm\mathbf{z}_{e}\left(\boldsymbol{\theta}\right)\mathbf{z}_{b}\herm\left(\boldsymbol{\theta}\right)\mathbf{z}_{b}\left(\boldsymbol{\theta}\right)}{1+P\mathbf{z}_{e}\left(\boldsymbol{\theta}\right)\mathbf{z}_{e}\herm\left(\boldsymbol{\theta}\right)}.
\end{align*}
Note that the terms $\mathbf{z}_{e}\left(\boldsymbol{\theta}\right)\mathbf{z}_{e}\herm\left(\boldsymbol{\theta}\right)$
in the denominator and $\mathbf{z}_{e}\left(\boldsymbol{\theta}\right)\mathbf{z}_{b}\herm\left(\boldsymbol{\theta}\right)$
in the numerator are scalars; both require $N$ complex multiplications,
whereas the term $\mathbf{z}_{e}\left(\boldsymbol{\theta}\right)\herm\mathbf{z}_{b}\left(\boldsymbol{\theta}\right)$
in the numerator needs $\mathcal{O}\left(N^{2}\right)$ complex multiplications.
Therefore, the complexity to compute $\left(P\mathbf{Z}_{e}\left(\boldsymbol{\theta}\right)+\mathbf{I}\right)^{-1}\left(P\mathbf{Z}_{b}\left(\boldsymbol{\theta}\right)+\mathbf{I}\right)$
is $\mathcal{O}\left(N^{2}\right)$. Reducing the complexity for computing
$\left(P\mathbf{Z}_{e}\left(\boldsymbol{\theta}\right)+\mathbf{I}\right)^{-1}\left(P\mathbf{Z}_{b}\left(\boldsymbol{\theta}\right)+\mathbf{I}\right)$
from $\mathcal{O}\left(N^{3}\right)$ to $\mathcal{O}\left(N^{2}\right)$
is particularly helpful for the case of extra-large MISO system where
the number of transmit antennas at Alice is very large. Moreover,
to find $\mathbf{u}_{\max}$, we need $\mathcal{O}\left(N^{3}\right)$
complex multiplications. Therefore, the overall complexity for each
update of $\mathbf{w}_{n}$ is given by $\mathcal{O}\left(NN_{\mathrm{ris}}+\tfrac{N^{2}}{2}+N^{2}+N^{3}\right)=\mathcal{O}\left(N^{3}+NN_{\mathrm{ris}}\right)$.
Following a similar line of argument, it can be noted that the number
of multiplications required to compute $\alpha_{bl}$, $\beta_{bl}$,
$\alpha_{el}$, and $\beta_{el}$ are all $\mathcal{O}\left(N_{\mathrm{ris}}\right)$.
Therefore, the complexity associated to update all phase shifts is
given by $\mathcal{O}\left(N_{\mathrm{ris}}^{2}\right)$. Hence, it
can be concluded that the overall complexity associated with each
iteration of Algorithm~\ref{alg:AO} (see lines 3\textendash 8) is
given by $\mathcal{O}\left(N^{3}+NN_{\mathrm{ris}}+N_{\mathrm{ris}}^{2}\right)$.

\section{Simulation Results and Discussion}

\begin{figure}[t]
\begin{centering}
\input{Fig_ConvergenceFINAL.tex}
\par\end{centering}
\caption{Convergence of ESR for $\vartheta\to0$ and $d_{ab,h}=10$ m.}
\label{fig:Convergence}
\end{figure}
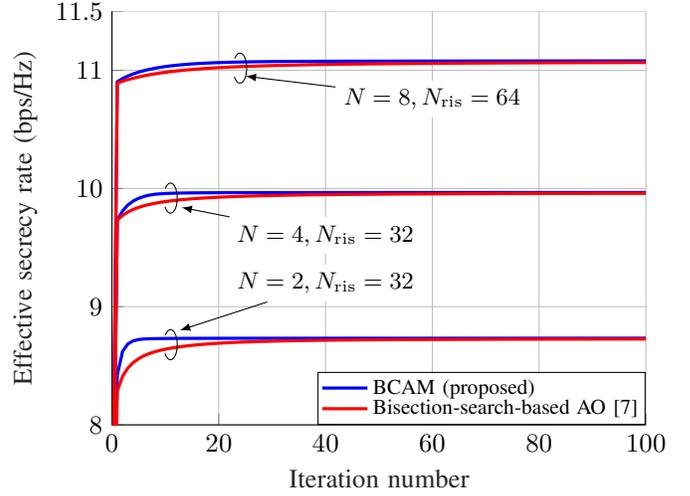

\begin{figure}[t]
\begin{centering}
\input{Fig_ESR_vs_Nris_VariableDelay.tex}
\par\end{centering}
\caption{The variation of ESR versus $N_{\mathrm{ris}}$ for $N=4$, $d_{ab,h}=40$
m and different values of $\vartheta$.}
\label{fig:VariableNris}
\end{figure}
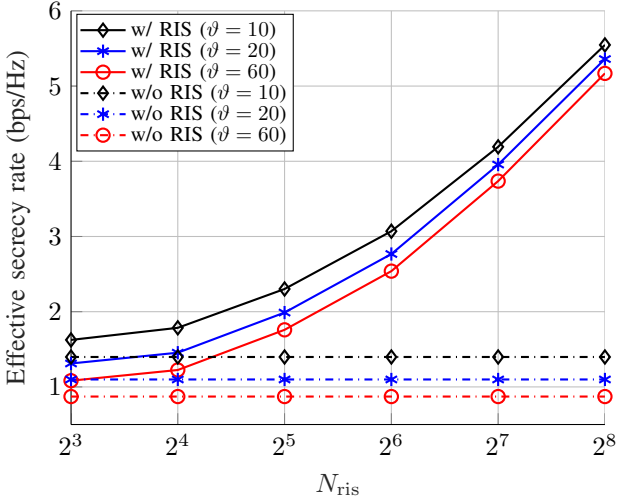

In this section, we present the simulation results for the ESR for
the MISO-RIS system. In order to facilitate a fair comparison with
a benchmark scheme, we consider the same system parameters as used
in~\cite{AO-Bisection}. For all of the wireless links, the small-scale
fading is modeled as Rayleigh distributed, whereas, the path loss
model is given by $\mathrm{PL}=\left[\mathrm{PL_{ref}}-10\xi\log_{10}\left(d/d_{\mathrm{ref}}\right)\right]$
dB. Here, $\mathrm{PL_{ref}}$ denotes the path loss at the reference
distance $d_{\mathrm{ref}}$, $\xi$ is the path loss exponent, and
$d$ is the distance between transmitter and receiver. The values
of different system parameters are given in~Table~\ref{Table-SysParameters}.
In Fig.~\ref{fig:Convergence}, the ESR is plotted for 50 different
channel realizations, whereas, in Figs.~\ref{fig:ESR_vsDistance}\textendash \ref{fig:VariableNris},
the ESR is plotted for $10^{3}$ different channel realizations.

\begin{table}[t]
\caption{System parameter values~\cite{AO-Bisection}.}
\label{Table-SysParameters}
\centering{}%
\begin{tabular}{|l|c|}
\hline 
\textbf{System Parameter} & \textbf{Value}\tabularnewline
\hline 
Transmit power, $P$ & 15 dBW\tabularnewline
\hline 
Noise power, $\sigma_{b}^{2}=\sigma_{e}^{2}$ & -75 dBW\tabularnewline
\hline 
Reference distance, $d_{\mathrm{ref}}$ & 1 m\tabularnewline
\hline 
Path loss at reference distance, $\mathrm{PL_{ref}}$ & -30 dBW\tabularnewline
\hline 
Path loss exponent for Alice-RIS links, $\xi_{ai}$ & 2.2\tabularnewline
\hline 
Path loss exponent for RIS-Bob links, $\xi_{ib}$ & 2.5\tabularnewline
\hline 
Path loss exponent for RIS-Eve links, $\xi_{ie}$ & 2.5\tabularnewline
\hline 
Path loss exponent for Alice-Bob links, $\xi_{ab}$ & 3.5\tabularnewline
\hline 
Path loss exponent for Alice-Eve links, $\xi_{ae}$ & 3.5\tabularnewline
\hline 
Alice-RIS distance, $d_{ai}$ & 50 m\tabularnewline
\hline 
Alice-Eve horizontal distance, $d_{ae,h}$ & 44 m\tabularnewline
\hline 
Alice-Bob horizontal distance (in meters) & $d_{ab,h}$\tabularnewline
\hline 
\tablefootnote{It is assumed that Bob and Eve lie in a horizontal line that is parallel
to that between Alice and the RIS.}Vertical distance between the line joining & \multirow{2}{*}{2 m}\tabularnewline
Alice and RIS, and Bob and Eve, i.e., $d_{v}$ & \tabularnewline
\hline 
\end{tabular}
\end{table}

Fig.~\ref{fig:Convergence} shows a comparison of the speed of convergence
between the proposed method (BCAM) and the bisection-search-based
AO given in~\cite{AO-Bisection}. Here one iteration constitutes
one update of the beamforming vector and one update of the phase shift
vector (see lines 3\textendash 8 in Algorithm~\ref{alg:AO}), which
is consistent with one iteration of the method in \cite{AO-Bisection},
and thus the comparison is fair. It is clear from the figure that
the proposed method converges significantly faster than that of the
algorithm adopted in~\cite{AO-Bisection}, establishing the superiority
of the proposed algorithm.

\begin{figure}[t]
\begin{centering}
\input{Fig_ESR_Comparison.tex}
\par\end{centering}
\caption{The variation in ESR versus the horizontal distance between Alice
and Bob for $N=4,N_{\mathrm{ris}}=32$ and different values of delay
QoS exponent.}
\label{fig:ESR_vsDistance}
\end{figure}
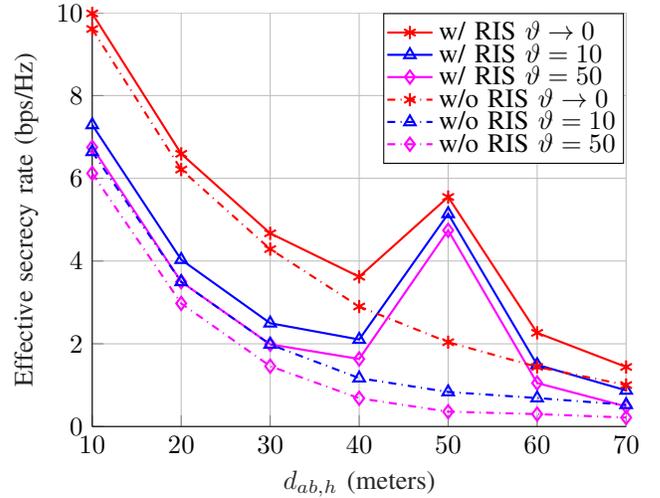

\begin{figure}[t]
\begin{centering}
\input{Fig_ESR_vs_Delay_VariableN.tex}
\par\end{centering}
\caption{The variation of ESR versus the delay exponent $\vartheta$ for $N_{\mathrm{ris}}=32$,
$d_{ab,h}=40$ m and different values of $N$.}
\label{fig:VariableN}
\end{figure}
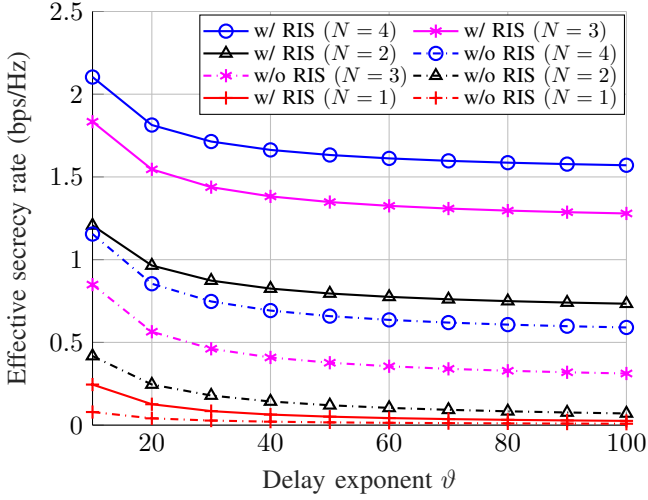

Fig.~\ref{fig:ESR_vsDistance} shows the variation in ESR versus
the horizontal distance between Alice and Bob (denoted by $d_{ab,h}$).
Note that for the system without any delay constraints (i.e., $\vartheta\to0$),
the ESR becomes equal to the ASR. It is evident from the figure that
as the delay requirement of the system becomes more stringent, i.e.,
for larger values of $\vartheta$, the ESR of the MISO-RIS system
decreases significantly. It can also be noted from the figure that
as the horizontal distance between Alice and Bob increases, the ESR
of the system decreases, because the Alice-Bob links becomes weak.
However, when $d_{ab,h}\in\left[40,50\right]$ m, the distance between
RIS and Bob becomes very small, and therefore, the RIS-Bob links become
very strong, resulting in an increased ESR. Moreover, the superiority
of introducing RIS is also clearly evident from the figure, as the
ones with RIS result in a significantly higher ESR than those without
RIS, for which the ESR decreases monotonically with increasing value
of $d_{ab,h}$.

In Fig.~\ref{fig:VariableN}, we show the effect of increasing delay
QoS exponent on the ESR for different number of transmit antennas.
It is clear from the figure that as the delay constraints for the
system becomes more strict, the ESR decreases. As expected, the RIS-assisted
system outperforms its counterpart without RIS. More interestingly,
for the case when Alice is equipped with a single transmit antenna
$\left(N=1\right)$, the system exhibits very poor ESR. However, as
the number of transmit antennas at Alice increases, sharp energy-focused
beamforming can performed at Alice to enhance the secrecy rate performance.
It is also noteworthy that increasing the value of $N$ results in
diminishing returns.

Fig.~\ref{fig:VariableNris} shows the variation in the ESR w.r.t.
$N_{\mathrm{ris}}$ for different values of the delay QoS exponent
$\vartheta$. The performance superiority of the RIS-assisted system
over the ones without RIS is clearly evident from the figure. More
interestingly, it can be noted from the figure that for an exponential
increase in the number of reflecting elements at RIS, i.e., $N_{\mathrm{ris}}$,
the ESR increases exponentially. Moreover, it is also interesting
to note that as the value of $N_{\mathrm{ris}}$ increases, the difference
between the ESR of the RIS-assisted system with different $\vartheta$
values decreases. This occurs due to the fact that for a very large
value of $N_{\mathrm{ris}}$, the fluctuation in the term $R_{s}$$\left(\mathbf{w},\boldsymbol{\theta}\right)$
in~\eqref{eq:ESR_Def} becomes very less, and therefore, the effect
of the delay QoS exponent $\vartheta$ on the ESR becomes negligible.
This result indicates that a large number of reflecting elements in
the RIS helps reducing the degradation in the ESR for delay-constrained
systems.

\section{Conclusion}

In this paper, we considered the problem of maximizing the secrecy
rate of a MISO-RIS system subject to the total transmit power constraint
and the delay-limited QoS constraint. We proposed a block coordinate
ascent method to find closed-form expressions for the beamformer and
the phase shift vector to maximize the objective. The convergence
superiority of the proposed solution over the bisection-search-based
AO is confirmed via Monte-Carlo simulation. The simulation results
confirmed that the secrecy rate of the system under stringent delay
requirements is significantly lower than the achievable secrecy rate,
however, a large-size RIS can greatly enhance the secrecy rate performance
of the system under delay constraints. Our results also confirm that
the ESR of the MISO-RIS system increases with an increase in the number
of transmit antennas and/or the number of reflecting elements at the
RIS.

\bibliographystyle{IEEEtran}
\bibliography{ref}

\end{document}

%% file: Fig_ConvergenceFINAL.tex
%
%
\pgfplotsset{width=0.98\columnwidth,height =0.8\columnwidth, compat=1.6}
\begin{tikzpicture}

\begin{axis}[%
xmin=0,
xmax=100,
xlabel style={font=\color{white!15!black}},
xlabel={Iteration number},
ymin=8,
ymax=11.5,
ytick={8,9,10,11,11.5},
ylabel style={font=\color{white!15!black}},
ylabel={Effective secrecy rate (bps/Hz)},
axis background/.style={fill=white},
axis x line*=bottom,
axis y line*=left,
xmajorgrids,
ymajorgrids,
legend style={{nodes={scale=0.8, transform shape}}, at={(0.385,0)},  anchor=south west, draw=black,fill=white,legend cell align=left,inner sep=1pt,row sep = -2pt},
]
\addplot [color=blue, line width=1.2pt]
  table[row sep=crcr]{%
0	6.50191588069861\\
1	10.9032208694115\\
2	10.9301414298107\\
3	10.9518692965018\\
4	10.9697578443014\\
5	10.9846653007618\\
6	10.9971103647809\\
7	11.0079153607858\\
8	11.0170495993842\\
9	11.0249012023361\\
10	11.0315840042795\\
11	11.0373238500928\\
12	11.0423346146496\\
13	11.0467130007419\\
14	11.050545197053\\
15	11.0538570983716\\
16	11.0567532713324\\
17	11.0592655246174\\
18	11.0614371582201\\
19	11.0633442484724\\
20	11.0650255198135\\
21	11.0665007362811\\
22	11.0677913738365\\
23	11.0689299471819\\
24	11.0699347450153\\
25	11.0707993251196\\
26	11.0715521825449\\
27	11.0722192010942\\
28	11.0728029016239\\
29	11.0733140233607\\
30	11.0737681639385\\
31	11.0741729936511\\
32	11.0745358860433\\
33	11.0748597883728\\
34	11.0751521896778\\
35	11.0754127539893\\
36	11.0756447790758\\
37	11.0758542315839\\
38	11.0760442172316\\
39	11.0762173117822\\
40	11.0763761903667\\
41	11.0765230201614\\
42	11.0766592798834\\
43	11.076785946646\\
44	11.076904004195\\
45	11.0770143922269\\
46	11.0771179156793\\
47	11.0772151765034\\
48	11.0773062729479\\
49	11.0773913635615\\
50	11.0774717800864\\
51	11.0775486073725\\
52	11.0776223506459\\
53	11.0776933265086\\
54	11.0777617854826\\
55	11.0778278856018\\
56	11.0778916386533\\
57	11.0779529653567\\
58	11.0780117949663\\
59	11.0780681395107\\
60	11.0781221623111\\
61	11.0781742141829\\
62	11.0782243918855\\
63	11.078272729115\\
64	11.0783193827961\\
65	11.0783644995265\\
66	11.0784081852768\\
67	11.0784505307\\
68	11.0784916328725\\
69	11.0785315866584\\
70	11.0785704581774\\
71	11.0786083023223\\
72	11.0786451943597\\
73	11.0786812102126\\
74	11.078716403211\\
75	11.0787508104501\\
76	11.0787844645128\\
77	11.0788173908662\\
78	11.0788496010844\\
79	11.0788810930006\\
80	11.0789118560085\\
81	11.0789418772547\\
82	11.0789711482934\\
83	11.0789996697116\\
84	11.0790274496873\\
85	11.0790545017315\\
86	11.0790808462613\\
87	11.0791065125431\\
88	11.0791315385038\\
89	11.079155969696\\
90	11.0791798594909\\
91	11.0792032705618\\
92	11.0792262744036\\
93	11.0792489430456\\
94	11.0792713292417\\
95	11.0792934503249\\
96	11.0793153077875\\
97	11.0793369187722\\
98	11.0793583127669\\
99	11.0793795148242\\
100	11.0794005396946\\
};
\addlegendentry{BCAM (proposed)}

\addplot [color=red, line width=1.2pt]
  table[row sep=crcr]{%
0	6.50191588069861\\
1	10.8929277819255\\
2	10.9077908701196\\
3	10.9211152917131\\
4	10.9330410798056\\
5	10.943719039503\\
6	10.9532956772291\\
7	10.9619024703097\\
8	10.9696534556461\\
9	10.9766471154575\\
10	10.982969128432\\
11	10.9886946448473\\
12	10.9938899046663\\
13	10.9986133973363\\
14	11.0029167608481\\
15	11.0068455440491\\
16	11.0104398944239\\
17	11.0137351883405\\
18	11.0167626075745\\
19	11.0195496581445\\
20	11.0221206355637\\
21	11.0244970285927\\
22	11.0266978774978\\
23	11.028740081062\\
24	11.0306386628592\\
25	11.0324070008665\\
26	11.0340570262586\\
27	11.0355993934447\\
28	11.0370436305528\\
29	11.0383982688913\\
30	11.0396709548329\\
31	11.0408685510904\\
32	11.0419972231789\\
33	11.0430625169449\\
34	11.0440694266814\\
35	11.045022455987\\
36	11.0459256708384\\
37	11.0467827463611\\
38	11.0475970103447\\
39	11.048371479538\\
40	11.0491088934876\\
41	11.0498117433536\\
42	11.050482298942\\
43	11.0511226307169\\
44	11.0517346316062\\
45	11.0523200344489\\
46	11.0528804286125\\
47	11.0534172741545\\
48	11.0539319152731\\
49	11.0544255915279\\
50	11.0548994477177\\
51	11.0553545435582\\
52	11.0557918617219\\
53	11.0562123145241\\
54	11.0566167516318\\
55	11.0570059644637\\
56	11.0573806927689\\
57	11.057741628431\\
58	11.0580894200823\\
59	11.0584246766766\\
60	11.058747970996\\
61	11.0590598425628\\
62	11.0593608006517\\
63	11.0596513258464\\
64	11.05993187332\\
65	11.0602028741136\\
66	11.060464736732\\
67	11.060717849056\\
68	11.0609625797824\\
69	11.0611992790365\\
70	11.0614282799925\\
71	11.0616499001447\\
72	11.0618644418046\\
73	11.0620721928127\\
74	11.0622734278425\\
75	11.0624684089636\\
76	11.0626573856778\\
77	11.0628405963205\\
78	11.0630182681434\\
79	11.0631906182224\\
80	11.0633578537182\\
81	11.0635201721908\\
82	11.0636777624246\\
83	11.0638308048211\\
84	11.0639794716474\\
85	11.0641239273249\\
86	11.0642643290717\\
87	11.0644008270601\\
88	11.0645335648627\\
89	11.0646626796763\\
90	11.0647883026403\\
91	11.0649105591529\\
92	11.0650295693788\\
93	11.0651454482876\\
94	11.0652583056332\\
95	11.06536824692\\
96	11.0654753728107\\
97	11.0655797800264\\
98	11.0656815609882\\
99	11.0657808043323\\
100	11.0658775953584\\
};
\addlegendentry{Bisection-search-based AO~\cite{AO-Bisection}}
\node[align=center,fill=white,inner sep=3pt] (D1) at (axis cs: 60, 10.76) {\small{$N = 8, N_{\mathrm{ris}} = 64$}};
\draw (axis cs: 23, 11.1) arc (140:-145: 0.09cm and  0.20cm);
\draw[->,>=latex] (D1) -- (axis cs:25, 10.95);

\addplot [color=blue, line width=1.2pt, forget plot]
  table[row sep=crcr]{%
0	5.95050428175197\\
1	8.42206379252041\\
2	8.62300979271261\\
3	8.68589326345385\\
4	8.71102703920479\\
5	8.72208520466519\\
6	8.72670226989953\\
7	8.72863495025903\\
8	8.72965115840808\\
9	8.73030442312319\\
10	8.73080264958342\\
11	8.73124583709441\\
12	8.73154748682286\\
13	8.73176336878916\\
14	8.73192882862954\\
15	8.73206004523596\\
16	8.73216475979893\\
17	8.73224920812834\\
18	8.73231863001558\\
19	8.73237675805581\\
20	8.73242643516651\\
21	8.73246986129654\\
22	8.73250861151066\\
23	8.73254363761694\\
24	8.73257560373375\\
25	8.73260507074915\\
26	8.73263258653499\\
27	8.73265857303939\\
28	8.73268313373275\\
29	8.73270622861594\\
30	8.73272786516324\\
31	8.73274807330414\\
32	8.73276690222651\\
33	8.73278443877587\\
34	8.73280081624404\\
35	8.73281619594437\\
36	8.7328307322654\\
37	8.73284455351749\\
38	8.7328577629794\\
39	8.73287044386608\\
40	8.73288266281183\\
41	8.7328944703895\\
42	8.73290589683285\\
43	8.73291694715839\\
44	8.73292760269484\\
45	8.73293783088636\\
46	8.7329475981754\\
47	8.73295687943729\\
48	8.73296566178803\\
49	8.73297394469978\\
50	8.73298173889089\\
51	8.73298906504878\\
52	8.73299595243622\\
53	8.73300243724691\\
54	8.73300856070865\\
55	8.73301436707253\\
56	8.7330199017197\\
57	8.73302520964146\\
58	8.73303033448047\\
59	8.73303531816096\\
60	8.73304020094086\\
61	8.7330450215915\\
62	8.73304981741752\\
63	8.73305462393547\\
64	8.73305947415831\\
65	8.7330643975371\\
66	8.73306941868577\\
67	8.7330745560665\\
68	8.73307982084715\\
69	8.73308521615868\\
70	8.73309073696263\\
71	8.7330963706616\\
72	8.73310209845153\\
73	8.73310789726824\\
74	8.73311374205999\\
75	8.73311960801122\\
76	8.73312547228337\\
77	8.73313131495319\\
78	8.73313711914419\\
79	8.73314287065185\\
80	8.73314855741446\\
81	8.73315416901077\\
82	8.73315969620053\\
83	8.73316513051036\\
84	8.73317046394324\\
85	8.73317568891905\\
86	8.73318079847043\\
87	8.7331857865758\\
88	8.73319064841288\\
89	8.73319538032857\\
90	8.73319997943173\\
91	8.73320444286371\\
92	8.73320876693526\\
93	8.73321294642625\\
94	8.7332169744224\\
95	8.73322084302821\\
96	8.73322454500694\\
97	8.73322807585246\\
98	8.73323143537381\\
99	8.73323462807503\\
100	8.73323766237114\\
};
\addplot [color=red, line width=1.2pt, forget plot]
  table[row sep=crcr]{%
0	5.95050428175197\\
1	8.28397901271963\\
2	8.412196487224\\
3	8.48180396064394\\
4	8.52749505325864\\
5	8.56002404682703\\
6	8.58439578656145\\
7	8.60326982654483\\
8	8.6182123627497\\
9	8.63028378183745\\
10	8.64025170070002\\
11	8.64866686895552\\
12	8.65591340826631\\
13	8.66225083064929\\
14	8.66784987764305\\
15	8.67282473445484\\
16	8.67725998538227\\
17	8.68122642397586\\
18	8.68478512187902\\
19	8.68798665086827\\
20	8.69087129354569\\
21	8.69347098082634\\
22	8.69581215106231\\
23	8.69791915478048\\
24	8.69981697220631\\
25	8.70153172567829\\
26	8.70308912829729\\
27	8.70451246222907\\
28	8.70582137562224\\
29	8.70703172690283\\
30	8.70815607397451\\
31	8.70920435205417\\
32	8.71018447725304\\
33	8.71110279257763\\
34	8.71196438053982\\
35	8.71277329873493\\
36	8.71353278798037\\
37	8.71424548076261\\
38	8.71491361161842\\
39	8.71553921192818\\
40	8.7161242696273\\
41	8.71667083348943\\
42	8.71718105870018\\
43	8.71765720091226\\
44	8.71810157702213\\
45	8.71851650928203\\
46	8.71890426807855\\
47	8.71926702510114\\
48	8.71960681991247\\
49	8.71992554053551\\
50	8.72022491743647\\
51	8.72050652516268\\
52	8.72077179230097\\
53	8.72102201372833\\
54	8.72125836449755\\
55	8.72148191383782\\
56	8.72169363814768\\
57	8.7218944320219\\
58	8.72208511759341\\
59	8.72226645242141\\
60	8.7224391357951\\
61	8.72260381355923\\
62	8.72276108228129\\
63	8.72291149278032\\
64	8.72305555323209\\
65	8.72319373192357\\
66	8.7233264597045\\
67	8.72345413220761\\
68	8.7235771124339\\
69	8.72369573283421\\
70	8.72381029787685\\
71	8.72392108558113\\
72	8.72402834979184\\
73	8.72413232178259\\
74	8.72423321233392\\
75	8.72433121341135\\
76	8.72442649941663\\
77	8.72451922891365\\
78	8.72460954603142\\
79	8.72469758156002\\
80	8.72478345417357\\
81	8.72486727163601\\
82	8.72494913175387\\
83	8.72502912326075\\
84	8.72510732685954\\
85	8.7251838157733\\
86	8.7252586565464\\
87	8.72533191013786\\
88	8.72540363196725\\
89	8.72547387269888\\
90	8.7255426787454\\
91	8.72561009266034\\
92	8.72567615364886\\
93	8.72574089792546\\
94	8.72580435887738\\
95	8.72586656748344\\
96	8.72592755238212\\
97	8.72598734032828\\
98	8.72604595620957\\
99	8.72610342330092\\
100	8.72615976353391\\
};
\addplot [color=blue, line width=1.2pt, forget plot]
  table[row sep=crcr]{%
0	6.1919060052333\\
1	9.730406714298\\
2	9.81068268941333\\
3	9.86095396208625\\
4	9.89432256928814\\
5	9.91698159374104\\
6	9.93277692218513\\
7	9.94376777998634\\
8	9.9508665034563\\
9	9.95487903370753\\
10	9.95746898322265\\
11	9.95936212384155\\
12	9.96066125413415\\
13	9.96154633473202\\
14	9.96220529198889\\
15	9.96272741362012\\
16	9.96315806728994\\
17	9.96351067957822\\
18	9.96379839801295\\
19	9.96403714041905\\
20	9.96423804452658\\
21	9.96440272240802\\
22	9.96454100111331\\
23	9.96466331534501\\
24	9.96477260995103\\
25	9.96486804955611\\
26	9.96495056757161\\
27	9.96502245831129\\
28	9.96508614769632\\
29	9.96514341192578\\
30	9.96519537214437\\
31	9.96524252576975\\
32	9.96528528477355\\
33	9.96532411770471\\
34	9.96535938000372\\
35	9.96539150553203\\
36	9.96542106028063\\
37	9.965448621911\\
38	9.96547471268451\\
39	9.96549981018602\\
40	9.96552425580835\\
41	9.96554809915867\\
42	9.9655711084087\\
43	9.96559298185384\\
44	9.96561355901805\\
45	9.96563285511863\\
46	9.96565098609406\\
47	9.96566809335657\\
48	9.965684301634\\
49	9.96569970620797\\
50	9.96571438148529\\
51	9.9657283947881\\
52	9.96574181159264\\
53	9.96575469661774\\
54	9.96576710995746\\
55	9.96577909829241\\
56	9.96579069748774\\
57	9.96580194230351\\
58	9.96581286895007\\
59	9.96582351114524\\
60	9.96583389565265\\
61	9.96584404001641\\
62	9.96585395310331\\
63	9.96586363785242\\
64	9.9658730943082\\
65	9.96588232148548\\
66	9.96589131829753\\
67	9.96590008438764\\
68	9.96590862097276\\
69	9.96591693115086\\
70	9.9659250194302\\
71	9.96593289100539\\
72	9.96594055154171\\
73	9.96594800767699\\
74	9.96595526775737\\
75	9.96596234215979\\
76	9.96596924291653\\
77	9.96597598278309\\
78	9.96598257407108\\
79	9.96598902752461\\
80	9.96599535140955\\
81	9.96600155090675\\
82	9.96600762786826\\
83	9.96601358097921\\
84	9.96601940633671\\
85	9.96602509839695\\
86	9.96603065115713\\
87	9.96603605936257\\
88	9.96604131950543\\
89	9.96604643043643\\
90	9.96605139352689\\
91	9.96605621244205\\
92	9.96606089266166\\
93	9.96606544089749\\
94	9.9660698645254\\
95	9.9660741711027\\
96	9.96607836800096\\
97	9.96608246215669\\
98	9.96608645992833\\
99	9.96609036704125\\
100	9.96609418860154\\
};
\addplot [color=red, line width=1.2pt, forget plot]
  table[row sep=crcr]{%
0	6.1919060052333\\
1	9.73269075619463\\
2	9.77227637357613\\
3	9.80134462347244\\
4	9.82326862933195\\
5	9.84031513149447\\
6	9.85396430272093\\
7	9.86515577125394\\
8	9.87450348842138\\
9	9.88243014533102\\
10	9.8892407045598\\
11	9.89516052609768\\
12	9.9003570019373\\
13	9.90495470283488\\
14	9.90904756179402\\
15	9.91270853384926\\
16	9.91599645293133\\
17	9.91896028471355\\
18	9.92164152270896\\
19	9.92407556069468\\
20	9.92629262035495\\
21	9.92831851192185\\
22	9.93017530608965\\
23	9.93188191791014\\
24	9.93345459977189\\
25	9.93490734358069\\
26	9.93625221332624\\
27	9.93749962844262\\
28	9.93865861383766\\
29	9.93973702364796\\
30	9.9407417471276\\
31	9.94167888774636\\
32	9.94255391587289\\
33	9.94337179134799\\
34	9.94413705254524\\
35	9.94485388122426\\
36	9.94552614193801\\
37	9.94615740813966\\
38	9.94675097795008\\
39	9.94730988523716\\
40	9.94783691112442\\
41	9.94833459478928\\
42	9.94880524683707\\
43	9.94925096506481\\
44	9.94967365049604\\
45	9.95007502476156\\
46	9.95045664870605\\
47	9.95081993968149\\
48	9.95116618857378\\
49	9.95149657576661\\
50	9.95181218605821\\
51	9.95211402140471\\
52	9.95240301217475\\
53	9.95268002652237\\
54	9.95294587821147\\
55	9.95320133161824\\
56	9.95344710578785\\
57	9.95368387622478\\
58	9.95391227510075\\
59	9.95413288888735\\
60	9.95434625621653\\
61	9.95455286472418\\
62	9.95475314640841\\
63	9.954947474793\\
64	9.95513616316496\\
65	9.95531946433424\\
66	9.95549757361922\\
67	9.95567063422123\\
68	9.9558387459992\\
69	9.95600197652739\\
70	9.95616037249344\\
71	9.9563139726558\\
72	9.95646281846436\\
73	9.95660696327792\\
74	9.95674647878362\\
75	9.95688145804541\\
76	9.95701201564989\\
77	9.95713828642736\\
78	9.95726042170601\\
79	9.95737858497955\\
80	9.95749294667534\\
81	9.95760368049932\\
82	9.95771095921085\\
83	9.95781495137543\\
84	9.9579158194121\\
85	9.95801371758127\\
86	9.95810879164097\\
87	9.95820117819529\\
88	9.95829100504269\\
89	9.95837839125614\\
90	9.95846344827576\\
91	9.95854628033735\\
92	9.95862698492564\\
93	9.95870565393135\\
94	9.95878237389956\\
95	9.95885722660624\\
96	9.95893028989596\\
97	9.9590016377864\\
98	9.95907134068682\\
99	9.95913946617329\\
100	9.95920607851133\\
};
\node[align=center,fill=white,inner sep=3pt] (D1) at (axis cs: 40, 9.6) {\small{$N = 4, N_{\mathrm{ris}} = 32$}};
\draw (axis cs:10, 10) arc (140:-145: 0.09cm and  0.20cm);
\draw[->,>=latex] (D1) -- (axis cs:13, 9.84);

\node[align=center,fill=white,inner sep=3pt] (D1) at (axis cs: 40, 9.2) {\small{$N = 2, N_{\mathrm{ris}} = 32$}};
\draw (axis cs:10, 8.76) arc (140:-145: 0.09cm and  0.20cm);
\draw[->,>=latex] (D1) -- (axis cs:12.7, 8.8);

\end{axis}
\end{tikzpicture}%

%% file: Fig_ESR_Comparison.tex
%
%
\pgfplotsset{width=0.98\columnwidth,height =0.8\columnwidth, compat=1.6}
\definecolor{mycolor1}{rgb}{1.00000,0.00000,1.00000}%
\begin{tikzpicture}

\begin{axis}[%
xmin=10,
xmax=70,
xlabel style={font=\color{white!15!black}},
xlabel={$d_{ab, h}$ (meters)},
ymin=0,
ymax=10,
ylabel style={font=\color{white!15!black}},
ylabel={Effective secrecy rate (bps/Hz)},
axis background/.style={fill=white},
axis x line*=bottom,
axis y line*=left,
xmajorgrids,
ymajorgrids,
legend style={at={(0.545,0.635)}, anchor=south west, draw=black,fill=white,legend cell align=left,inner sep=1pt,row sep = -4pt},
]
\addplot [color=red, mark=asterisk, line width=0.8pt, mark size=2.5pt, mark options={solid, red}]
  table[row sep=crcr]{%
10	9.9858\\
20	6.5918\\
30	4.6704\\
40	3.62\\
50	5.548\\
60	2.2645\\
70	1.4391\\
};
\addlegendentry{w/ RIS $\vartheta \rightarrow 0$}


\addplot [color=blue, mark=triangle, line width=0.8pt, mark size=2.5pt, mark options={solid, blue}]
  table[row sep=crcr]{%
10	7.291\\
20	4.0339\\
30	2.4971\\
40	2.1036\\
50	5.1409\\
60	1.4923\\
70	0.8728\\
};
\addlegendentry{w/ RIS $\vartheta = 10$}

\addplot [color=mycolor1, mark=diamond, line width=0.8pt, mark size=2.5pt, mark options={solid, mycolor1}]
  table[row sep=crcr]{%
10	6.7505\\
20	3.4991\\
30	1.9864\\
40	1.6326\\
50	4.7421\\
60	1.051\\
70	0.4845\\
};
\addlegendentry{w/ RIS $\vartheta = 50$}

\addplot [color=red, dashdotted, line width=0.8pt, mark size=2.5pt, mark=asterisk, mark options={solid, red}]
  table[row sep=crcr]{%
10	9.60906929824188\\
20	6.21085904578034\\
30	4.28955292213961\\
40	2.90040347557879\\
50	2.03795362414393\\
60	1.44163129930424\\
70	1.00260445039413\\
};
\addlegendentry{w/o RIS $\vartheta \rightarrow 0$}

\addplot [color=blue, dashdotted, line width=0.8pt, mark size=2.5pt, mark=triangle, mark options={solid, blue}]
  table[row sep=crcr]{%
10	6.64107059217141\\
20	3.49672687864998\\
30	1.98339232527262\\
40	1.16921916807879\\
50	0.835253229275308\\
60	0.688140216155109\\
70	0.525542489445752\\
};
\addlegendentry{w/o RIS $\vartheta = 10$}

\addplot [color=mycolor1, dashdotted, line width=0.8pt, mark size=2.5pt, mark=diamond, mark options={solid, mycolor1}]
  table[row sep=crcr]{%
10	6.12014657527645\\
20	2.97562110902354\\
30	1.45877036261225\\
40	0.682829610132972\\
50	0.358865157635782\\
60	0.299967176396631\\
70	0.216943442900062\\
};
\addlegendentry{w/o RIS  $\vartheta = 50$}

\end{axis}
\end{tikzpicture}%

%% file: Fig_ESR_vs_Delay_VariableN.tex
%
%
\pgfplotsset{width=0.98\columnwidth,height =0.8\columnwidth, compat=1.6}
\definecolor{mycolor1}{rgb}{1.00000,0.00000,1.00000}%
\begin{tikzpicture}

\begin{axis}[%
xmin=10,
xmax=100,
xlabel style={font=\color{white!15!black}},
xlabel={Delay exponent $\vartheta$},
ymin=0,
ymax=2.5,
ylabel style={font=\color{white!15!black}},
ylabel={Effective secrecy rate (bps/Hz)},
axis background/.style={fill=white},
axis x line*=bottom,
axis y line*=left,
xmajorgrids,
ymajorgrids,
legend columns=2, 
legend style={{nodes={scale=0.8, transform shape}}, at={(0.20,0.75)},  anchor=south west, draw=black,fill=white,legend cell align=left,inner sep=1pt,row sep = -2pt},
]
\addplot [color=blue, line width=0.8pt, mark=o, mark size=2.5pt, mark options={solid, blue}]
  table[row sep=crcr]{%
10	2.10362683177664\\
20	1.81367433183475\\
30	1.71387425907732\\
40	1.6632454017259\\
50	1.63257165020499\\
60	1.6119946079638\\
70	1.59723912738345\\
80	1.58614485117407\\
90	1.57750135666982\\
100	1.570577828628\\
};
\addlegendentry{w/ RIS $(N = 4)$}

\addplot [color=mycolor1, line width=0.8pt,  mark size=2.5pt, mark=asterisk, mark options={solid, mycolor1}]
  table[row sep=crcr]{%
10	1.8335099636488\\
20	1.54589202709819\\
30	1.43821390891804\\
40	1.38226979194956\\
50	1.34815829838273\\
60	1.32525777906738\\
70	1.30884996639014\\
80	1.29652748500114\\
90	1.28693767374434\\
100	1.2792638597409\\
};
\addlegendentry{w/ RIS $(N = 3)$}

\addplot [color=black, line width=0.8pt,  mark size=2.5pt, mark=triangle, mark options={solid, black}]
  table[row sep=crcr]{%
10	1.2069775474717\\
20	0.96457173868594\\
30	0.873391881739443\\
40	0.82527397813479\\
50	0.795461067138747\\
60	0.775188983301552\\
70	0.760523356187039\\
80	0.749426876629446\\
90	0.740739153163759\\
100	0.733751749761461\\
};
\addlegendentry{w/ RIS $(N = 2)$}

\addplot [color=blue, dashdotted, line width=0.8pt, mark=o,  mark size=2.5pt, mark options={solid, blue}]
  table[row sep=crcr]{%
10	1.15490179429017\\
20	0.854489883998226\\
30	0.747371033763696\\
40	0.692306820859861\\
50	0.658692108917342\\
60	0.636033681682395\\
70	0.619739010920857\\
80	0.60746876115932\\
90	0.597903012901563\\
100	0.590240304280788\\
};
\addlegendentry{w/o RIS $(N = 4)$}

\addplot [color=mycolor1, dashdotted, line width=0.8pt, mark=asterisk, mark size=2.5pt, mark options={solid, mycolor1}]
  table[row sep=crcr]{%
10	0.849104768724633\\
20	0.564554725813428\\
30	0.461380518774046\\
40	0.408911818180613\\
50	0.377101540459115\\
60	0.355690639881248\\
70	0.340253641457251\\
80	0.328572331690606\\
90	0.319411804064744\\
100	0.312028976990166\\
};
\addlegendentry{w/o RIS $(N = 3)$}

\addplot [color=black, dashdotted, line width=0.8pt, mark=triangle,  mark size=2.5pt, mark options={solid, black}]
  table[row sep=crcr]{%
10	0.417453948934639\\
20	0.245479723932815\\
30	0.178925904592279\\
40	0.142887794554498\\
50	0.120061571800085\\
60	0.104217214258019\\
70	0.0925332927128474\\
80	0.0835366130081472\\
90	0.076379208524005\\
100	0.0705371650683764\\
};
\addlegendentry{w/o RIS $(N = 2)$}

\addplot [color=red, line width=0.8pt, mark=+,  mark size=2.5pt,  mark options={solid, red}]
  table[row sep=crcr]{%
10	0.24522331175771\\
20	0.126299924429186\\
30	0.0848295957877345\\
40	0.0638189591784067\\
50	0.0511359232816695\\
60	0.0426526803965214\\
70	0.0365811169842611\\
80	0.0320214160014078\\
90	0.028471649390676\\
100	0.0256298514628371\\
};
\addlegendentry{w/ RIS $(N = 1)$}

\addplot [color=red, dashdotted, line width=0.8pt,  mark size=2.5pt, mark=+, mark options={solid, red}]
  table[row sep=crcr]{%
10	0.0792020802182383\\
20	0.0410092564434944\\
30	0.0276469501300236\\
40	0.0208549697741595\\
50	0.0167442164682119\\
60	0.0139883863396151\\
70	0.0120121256811129\\
80	0.0105255122311701\\
90	0.00936657351372778\\
100	0.00843770265776312\\
};
\addlegendentry{w/o RIS $(N = 1)$}

\end{axis}
\end{tikzpicture}%

%% file: Fig_ESR_vs_Nris_VariableDelay.tex
%
%
\pgfplotsset{width=0.98\columnwidth,height =0.8\columnwidth, compat=1.6}
\begin{tikzpicture}

\begin{axis}[%
xmode=log,
log basis x={2},
xmin=8,
xmax=256,
xtick={8,16,32,64,128,256},
xlabel style={font=\color{white!15!black}},
xlabel={$N_{\mathrm{ris}}$},
ymin=0.5,
ymax=6,
ylabel style={font=\color{white!15!black}},
ylabel={Effective secrecy rate (bps/Hz)},
axis background/.style={fill=white},
axis x line*=bottom,
axis y line*=left,
xmajorgrids,
ymajorgrids,
legend style={{nodes={scale=0.8, transform shape}}, at={(0.01,0.66)},  anchor=south west, draw=black,fill=white,legend cell align=left,inner sep=1pt,row sep = -2pt},
]
\addplot [color=black, line width=0.8pt, mark size=2.5pt, mark=diamond, mark options={solid, black}]
  table[row sep=crcr]{%
8	1.62489807168843 \\
16	1.78567092228949 \\
32	2.30189603446494 \\
64	3.070089192241 \\
128	4.19188416734432 \\
256	5.54696032905886 \\
};
\addlegendentry{w/ RIS ($\vartheta = 10$)}

\addplot [color=blue, line width=0.8pt, mark size=2.5pt, mark=asterisk, mark options={solid, blue}]
  table[row sep=crcr]{%
8	1.31144203971266\\
16	1.4547794416456\\
32	1.98840708994794\\
64	2.76695267498601\\
128	3.95615332977327\\
256	5.35935344895748\\
};
\addlegendentry{w/ RIS ($\vartheta = 20$)}

\addplot [color=red, line width=0.8pt, mark size=2.5pt, mark=o, mark options={solid, red}]
  table[row sep=crcr]{%
8	1.08263751628879\\
16	1.22483420284972\\
32	1.75901003116638\\
64	2.53933035913113\\
128	3.73628668848287\\
256	5.16845022572459\\
};
\addlegendentry{w/ RIS ($\vartheta = 60$)}

\addplot [color=black, dashdotted, line width=0.8pt, mark size=2.5pt, mark=diamond, mark options={solid, black}]
  table[row sep=crcr]{%
8	1.39789718498209\\
16	1.39789718498209\\
32	1.39789718498209\\
64	1.39789718498209\\
128	1.39789718498209\\
256	1.39789718498209\\
};
\addlegendentry{w/o RIS ($\vartheta = 10$)}

\addplot [color=blue, dashdotted, line width=0.8pt, mark size=2.5pt, mark=asterisk, mark options={solid, blue}]
  table[row sep=crcr]{%
8	1.09896227706414\\
16	1.09896227706414\\
32	1.09896227706414\\
64	1.09896227706414\\
128	1.09896227706414\\
256	1.09896227706414\\
};
\addlegendentry{w/o RIS ($\vartheta = 20$)}

\addplot [color=red, dashdotted, line width=0.8pt, mark size=2.5pt, mark=o, mark options={solid, red}]
  table[row sep=crcr]{%
8	0.871762559323347\\
16	0.871762559323347\\
32	0.871762559323347\\
64	0.871762559323347\\
128	0.871762559323347\\
256	0.871762559323347\\
};
\addlegendentry{w/o RIS ($\vartheta = 60$)}

\end{axis}
\end{tikzpicture}%